\documentstyle [12pt,epsf] {article}

\begin{document}
\begin{titlepage}
\begin{center}
\vspace{2cm}
\LARGE
Environmental Influences on Dark Matter Halos and Consequences for the Galaxies
Within Them
\\
\vspace{1cm}
\large
Gerard Lemson$^{1}$ \& Guinevere Kauffmann$^{2}$  \\
\vspace{0.5cm}
\small
{\em $^1$ Racah Institute of Physics, The Hebrew University, Jerusalem 91904,
Israel} \\
{\em $^2$Max-Planck Institut f\"{u}r Astrophysik, D-85740 Garching, Germany} \\
\vspace{0.8cm}
\end{center}
\normalsize
\begin {abstract}
We use large N-body simulations of dissipationless gravitational clustering
in cold dark matter (CDM) cosmologies
to study whether the properties of dark matter halos are affected by
their environment.
We look for correlations between the
masses, formation redshifts, concentrations, shapes and spins of halos and
the overdensity of their local environment.
We also look for correlations of these quantities with the local tidal field.
Our conclusion is extremely simple. Only the mass distribution varies
as a function of environment. This variation is well described
by a simple analytic formula based on the conditional Press-Schechter theory.
We find no significant dependence of any other halo property on environment.
Our results do not depend on our choice of cosmology.
According to current hierarchical models,  the structure
and evolutionary history of a galaxy is fully determined by the structure and
evolutionary
history of the dark halo in which it is embedded. If these models are correct,
clustering
variations between galaxies of differing morphological types, luminosities,
colours and
surface brightnesses, must all arise because the halo mass function is skewed
towards high
mass objects in overdense regions of the Universe and towards low mass objects
in underdense
regions.
\end {abstract}
\vspace {0.8 cm}
\end {titlepage}

\section {Introduction}
It is well known that many of the observed properties of galaxies correlate
with their
environment. The most famous of these correlations is the morphology-density
relation.
Davis \& Geller (1976) showed that elliptical galaxies are more strongly
clustered
on the sky than spirals, and Dressler (1980) demonstrated that the elliptical
fraction in
galaxy clusters is an increasing function of local density. The dependence of
the clustering
of galaxies on luminosity has been  much more controversial, but recent
analyses of the largest available redshift surveys confirm that
L$_{*}$ and brighter galaxies have a higher clustering amplitude than
low-luminosity galaxies
(Park et al 1994; Loveday et al 1995; Benoist et al 1996; Valotto \& Lambas
1997).
Low surface brightness galaxies are also distributed differently from their
high surface
brightness counterparts. They are less clustered on all scales (Mo, McGaugh \&
Bothun 1994)
and are particularly isolated on scales smaller than a few Mpc
(Bothun et al 1993). Finally, the star formation histories and the gas
fractions
of galaxies
of nominally the same Hubble type are influenced by their local environments.
Spiral galaxies in dense environments have redder colours and lower star
formation rates
that spirals in the field, and often also exhibit truncated HI disks
( e.g. Kennicutt 1983; Cayette et al 1994).

One popular hypothesis for the origin of clustering differences between
galaxies of different types
is that mergers, tidal encounters or interactions with a
surrounding gaseous medium modify galaxy properties. Such interactions are more
probable in high-density environments. It should be noted, however, that
differences
in the clustering amplitude of galaxies of different morphologies, luminosities
and surface
brightnesses persist out to scales where the crossing time is much larger
than the Hubble time (Mo et al 1992; Mo, McGaugh \& Bothun 1994), so it does
appear that
galaxies
bear some imprint of the initial field of density fluctuations
in the Universe.

According to the standard cosmological paradigm, structure in the Universe is
built up through
a process of hierarchical clustering. Small-scale fluctuations in the initial
density field
are the first to collapse to form bound, virialized objects (or {\em dark
matter halos})
and these merge together over time to form more and more massive
systems. Galaxies form when
gas cools, settles, and turns into stars at the centres of the halos (White \&
Rees 1978; White \&
Frenk 1991). Galaxy formation models of this type have met with considerable
success in explaining
many of the trends seen in the properties of galaxies,
both at present day and at high redshift
( e.g.  Lacey et al 1993; Kauffmann, White \& Guiderdoni 1993; Cole et al
1994).
Their most serious weakness arises from the fact that star formation and
supernova feedback processes
are poorly understood and hence difficult to model in a convincing way.

It should be noted, however, that one {\em fundamental} aspect of this picture
is that the structure and evolutionary history of a galaxy is {\em fully
determined} by
the structure and evolutionary history of its surrounding dark matter halo. The
merging
history of the halo determines the rate at which gas will cool and become
available for star formation, as well as the frequency of merging events and
the
mass distribution of accreted objects.
The density profile of the halo, its shape and
its distribution of angular momentum determine the structure, the size and the
rotation
curve of the galaxy that forms at its centre (Dalcanton, Spergel \&
Summers 1997; Mo, Mao \&
White 1997).
If the properties of galaxies are observed to vary as a function of
environment, it follows
that the properties of their surrounding halos must also vary with environment.
By understanding
exactly which halo properties can be affected by local conditions, one can hope
to gain a
deeper understanding of the origin of the clustering differences discussed
above.

In this paper, we use numerical simulations of gravitational clustering to
study
the properties of dark matter halos as a function of local density. In order to
achieve both
good statistics and an accurate treatment of the formation history and internal
structure
of halos, high resolution simulations are needed, which nevertheless contain a
fair sample
of the Universe, thus accounting correctly for the influence of large-scale
structure
on the formation of the halos. We look for correlations between a set of
present-day  halo properties,
including their masses, formation redshifts, concentrations, shapes and spins,
with
the overdensity of their local environment.
We also look for correlations of these quantities with the surrounding tidal
field.
Our conclusion is extremely simple. Only the mass distribution of dark halos
varies
as a function of environment. The variation in the mass function is described
extremely well by a simple analytic formula based on the conditional
Press-Schechter theory
derived by Mo \& White (1996). We find no significant dependence of any other
halo property
on environment.
Our results do not depend on our choice of cosmology.
This leads to the conclusion that the dependences of galaxy morphology,
luminosity, surface
brightness and star formation rate on environment, must {\em all}
arise because galaxies are
preferentially found in higher mass halos in overdense environments and in
lower mass halos in
underdense environments.
\section{The simulations}

The GIF project is a joint effort of astrophysicists from Germany and
Israel. Its primary goal is to study the formation and evolution of
galaxies in a cosmological context using semi-analytical galaxy
formation models embedded in large high-resolution $N$-body
simulations. This is done by constructing merger trees of particle
halos from dark-matter only simulations and placing galaxies into
them using a phenomenological modelling (for a detailed description of
this procedure as well as results cf.\ Kauffmann et al.\ 1997).

The code used for the GIF simulations is called Hydra. It is a
parallel adaptive particle-particle particle-mesh (AP$^3$M) code (for
details on the code cf.\ Couchman, Thomas, \& Pearce 1995; Pearce \&
Couchman 1997). The current version was developed as part of the VIRGO
supercomputing project and was kindly made available by them for the
GIF project. The simulations were started on the CRAY T3D at the
Computer Centre of the Max-Planck Society in Garching (RZG) on 128
processors. Once the clustering strength required an even larger
amount of total memory, they were transferred to the T3D at the
Edinburgh Parallel Computer Centre (EPCC) and finished on 256
processors.

A set of four simulations with $N=256^3$ and with different
cosmological parameters was run. In this paper we focus on a variant of a cold
dark
matter model, $\tau$CDM,
with $\Omega_0=1$, $h=0.5$, $\sigma_8=0.6$ and shape parameter $\Gamma=0.21$.
A value of $\Gamma=0.21$ is usually preferred by analyses of galaxy clustering,
cf.\ Peacock \& Dodds (1994). This is achieved in the $\tau$CDM model
despite $\Omega_0=1$ and $h=0.5$ by assuming that a massive neutrino
(usually taken to be the $\tau$ neutrino) had existed during the very
early evolution of the Universe and came to dominate the energy density for a
short period. It then decayed into lighter neutrinos which are still
relativistic, thus
delaying the epoch when matter again started to dominate over radiation.
The neutrino mass and lifetime are chosen such that
$\Gamma=0.21$. For a detailed description of such a model see White,
Gelmini, \& Silk (1995). The normalization was chosen so as to match the
abundance of rich clusters (White et al 1993). The simulation box size
was 85 h$^{-1}$ Mpc and the particle mass was $2 \times 10^{10} M_{\odot}$.
In order to be sure that our results
are not sensitive to our choice of cosmological initial conditions, we have
also
analyzed a low-density CDM model with $\Omega_0=0.3$, $\Omega_{\Lambda}=0.7$,
$\Gamma=0.21$,
h=0.7 and $\sigma_8=0.9$ ($\Lambda$CDM). This simulation has a somewhat larger
box size (141 h$^{-1}$ Mpc), but the same particle mass.

\section {Procedure}
Halos were selected from the $z=0$ output times in the simulations as follows.
First, we searched for high-density regions using a standard friends-of-friends
groupfinder
with a linking length of $b=0.2$ times the mean interparticle separation. We
then
searched for the particle with the lowest potential energy and
adopted its position as the halo centre. The distances of all the particles
to the
centre were ordered and the radius of the largest sphere with an overdensity
$\delta \ge 200$ was defined to be the {\em virial radius} of the halo, and the
mass
contained within this radius was defined to be the {\em virial mass}.
Halos were only
included in
our analysis if the friends-of-friends mass exceeded 70 particles and
the virial mass was between 35\% and 95\% of the
friends-of-friends mass. The upper bound served as a check that
the majority of particles at overdensities greater than $\simeq 200$ were
located
by the groupfinder. The lower bound was chosen to ensure that enough
particles were located within the virialized region so that the various
halo quantities could be calculated reliably.

The following quantities were evaluated for each halo:
\begin {enumerate}
\item The spin parameter $\lambda = L E^{1/2} /GM ^{5/2}$, calculated for
the particles within the virial radius, where $L$ and $E$ are the angular
momentum and the thermal kinetic
energy of the halo.
\item The formation redshift $z_{form}$, defined as the redshift at which the
most massive halo
progenitor was half the present-day mass of the halo.
\item Concentration indices $c_{10} = r_{10}/r_{vir}$  and $c_{20} =
 r_{20}/r_{vir}$,
 where $r_{10}$ and $r_{20}$ are the radii enclosing one tenth and one
twentieth of the
 virial mass respectively.
These parameters provide a coarse measure of the shape of the density profile
 of the halo.
\item The shape of the halo was determined by diagonalizing the moment of
 inertia tensor
 of halo particles within the virial radius. This gives the principal axes
$ p_1 \ge p_2 \ge p_3$.
\end {enumerate}

The properties of the local environment around each halo were determined as
follows.
We evaluated the overdensity $\delta$ in spheres of various radii
surrounding the halo. We also calculated overdensities in shells around the
halo,
thereby excluding the contribution of the halo mass itself.

In order to obtain a measure of the higher order harmonics of the surrounding
dark matter
distribution, a density field on a 128$^{3}$ grid
(cell size 0.66 h$^{-1}$ Mpc) was obtained from the simulation
 using cloud-in-cell (CIC) interpolation. The resulting
field was smoothed using a top-hat smoothing window of 10 h$^{-1}$ Mpc and  the
first and second-order
spatial derivatives of the density and potential fields were determined using
fast
Fourier transforms in k-space. We made use of the following k-space window
functions:
\begin {itemize}
\item density dipole: $ik_i$
\item density shear field: $-k_i k_j$
\item potential dipole field (acceleration): $-i k_i /k^2$
\item potential shear field (tidal field): $k_i k_j/ k^2$
\end {itemize}
The resulting fields were interpolated to the positions of the halos using the
CIC method.
The shear fields were diagonalized and the eigenvalues ordered.

\section {Results}
\subsection {The mass function of halos versus environment}
In an extension of the Press-Schechter theory, Bond
et al (1991) and Bower (1991)
derive an expression for the fraction of mass in a region of initial radius
$R_0$ and linear
overdensity $\delta_0$, which at redshift $z_1$ is contained in halos of mass
$M_1$.
If one assumes that the region $R_0$ evolves in size and overdensity according
to the spherical top-hat collapse model, it is simple to derive a formula for
the
mass function of halos in a present-day region of radius $R$ and overdensity
$\delta$. This was first done by Mo \& White (1996). They tested their analytic
formula against results from  N-body simulations
containing 128$^{3}$ particles.

In figure 1, we show how the shape of the halo mass function changes as a
function of
local overdensity $\delta$ evaluated within spheres of R= 10 h$^{-1}$ Mpc. The
thick
solid line shows results derived from the $\tau$CDM simulation at $z=0$.
The thin solid line shows the analytic prediction of Mo \& White (1996).
For comparison, the thick dashed line is the {\em average}
halo mass function evaluated from the simulation, multiplied by a factor
$(1+\delta)$. The thin dashed line is the same quantity
calculated from the Press-Schechter theory. For clarity, the simulation and
the analytic curves have been offset by 1 decade in the y-direction.

\begin{figure}
\centerline{
\epsfxsize=14cm \epsfbox{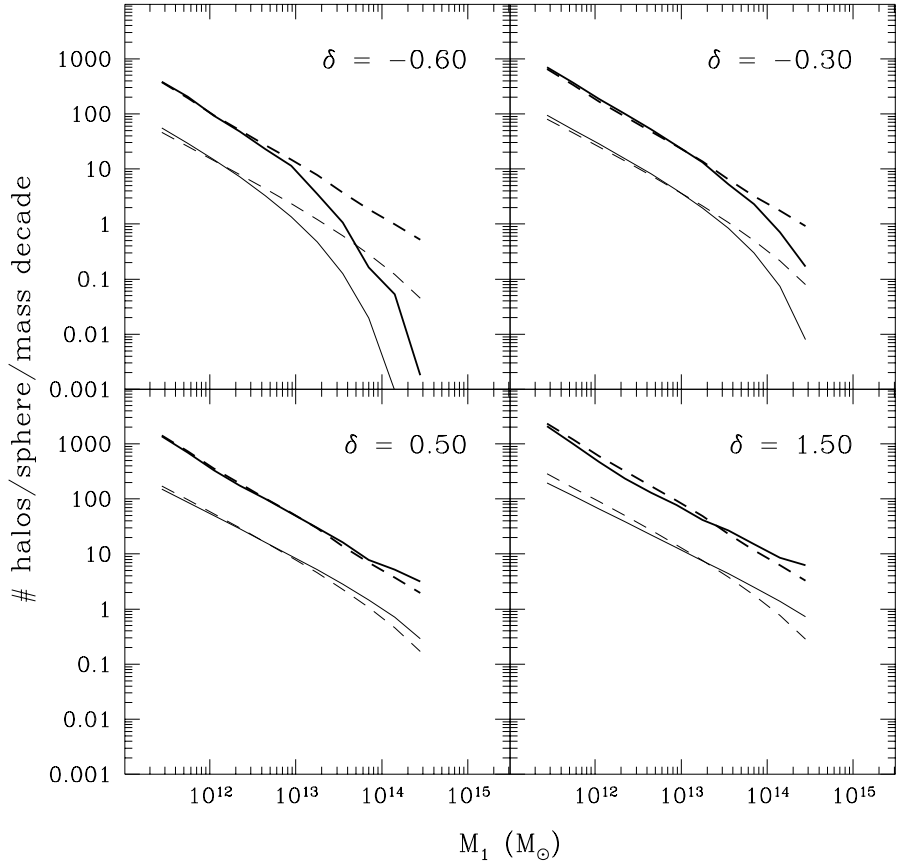}
}
\caption{\label{fig1}
The halo mass function in spheres of radius R=10 h$^{-1}$ Mpc and local
overdensity
$\delta$. The thick
solid line shows results derived from the $\tau$CDM simulation at $z=0$.
The thin solid line shows the analytic prediction of Mo \& White (1996).
For comparison, the thick dashed line is the {\em average}
halo mass function evaluated from the simulation, multiplied by a factor
$(1+\delta)$. The thin dashed line is the same quantity
calculated from the Press-Schechter theory. For clarity, the simulation and
the analytic curves have been offset by 1 decade in the y-direction.}
\end {figure}

Although it is not apparent as a result of the way we have chosen to
present our results, it should be noted that
the halo abundances predicted by the analytic theory exceed
those derived from the simulation by a factor $\simeq 1.5$ over the range of
halo masses shown
in figure 1.
The magnitude of this offset is in rough agreement with
that found
by Lacey \& Cole (1994) in their tests of the Press-Schechter formalism.
Nevertheless,
as can be seen in the plot, the theory
does predict the change in shape of the halo mass function with $\delta$
remarkably well.
As can be seen, in low density regions, high-mass halos are underepresented,
whereas in high-density regions the situation is reversed and
high-mass halos are overabundant.

\subsection {Formation times, Spins, Concentrations and Shapes of Halos}
Our principal result is that we find {\em no correlation} between the
environment
of halos and their formation times, spins,
concentrations and shapes in either the $\tau$CDM
or the $\Lambda$CDM simulations. Out of a large number of possible
scatterplots,
we have
selected the following few to illustrate this conclusion. Results are shown
only
for the $\tau$CDM model, but plots for $\Lambda$CDM are virtually identical.

In figure 2 we plot the value of the spin parameter $\lambda$ against the
overdensity in top-hat
spheres of radius 10 h$^{-1}$ Mpc for halos in four different mass bins.
This figure shows that the mean value of $\lambda$ varies at most very weakly
with overdensity or with mass. There does appear to be a slight tendency for
$\lambda$ to increase with overdensity. These results do not change if we adopt
a different value of the smoothing radius.

\begin{figure}
\centerline{
\epsfxsize=14cm \epsfbox{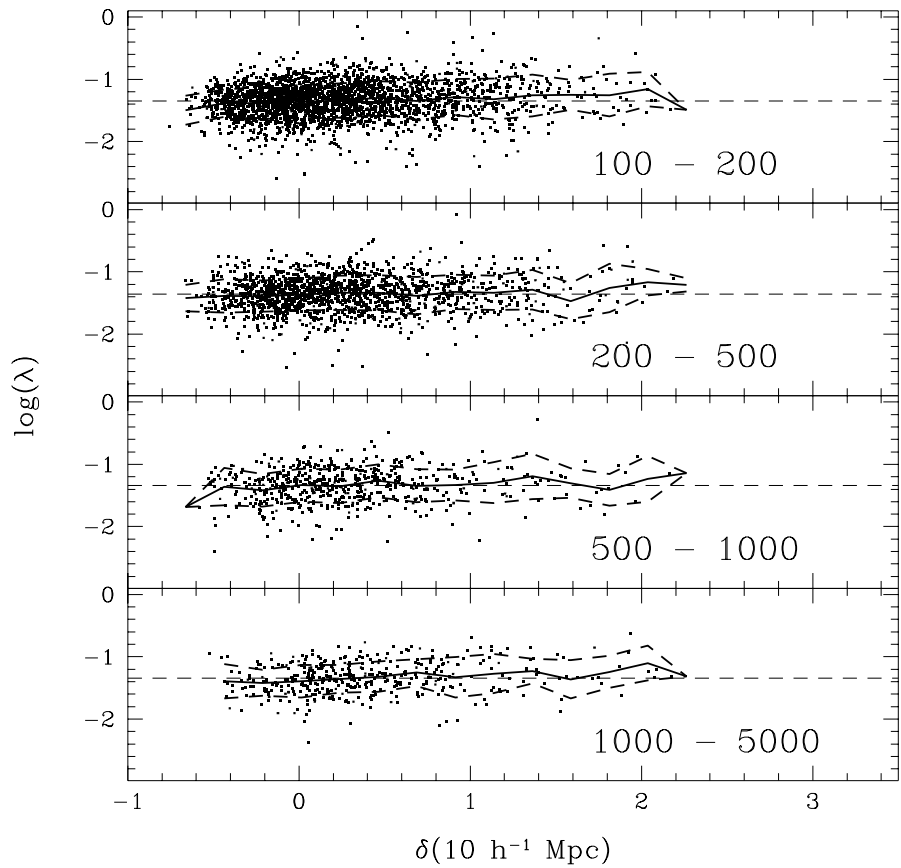}
}
\caption{\label{fig2}
The spin parameter $\lambda$ is plotted against overdensity in spheres of
radius R=10 h$^{-1}$ Mpc for
halos in different mass ranges, indicated on each panel as the number of
particles found within the
virial radius.
The solid line shows the mean value of $ \log (\lambda$) as a function of
$\delta$.
Dashed lines indicate the 1$\sigma$ standard deviation. }
\end {figure}

In figure 3 we plot formation redshift against the overdensity in a shell
between
2 h$^{-1}$
and 5 h$^{-1}$ Mpc surrounding each halo. No dependence of mean formation
redshift on
overdensity is seen. Higher mass halos form somewhat later on average, as
expected (Lacey \& Cole 1993).

\begin{figure}
\centerline{
\epsfxsize=14cm \epsfbox{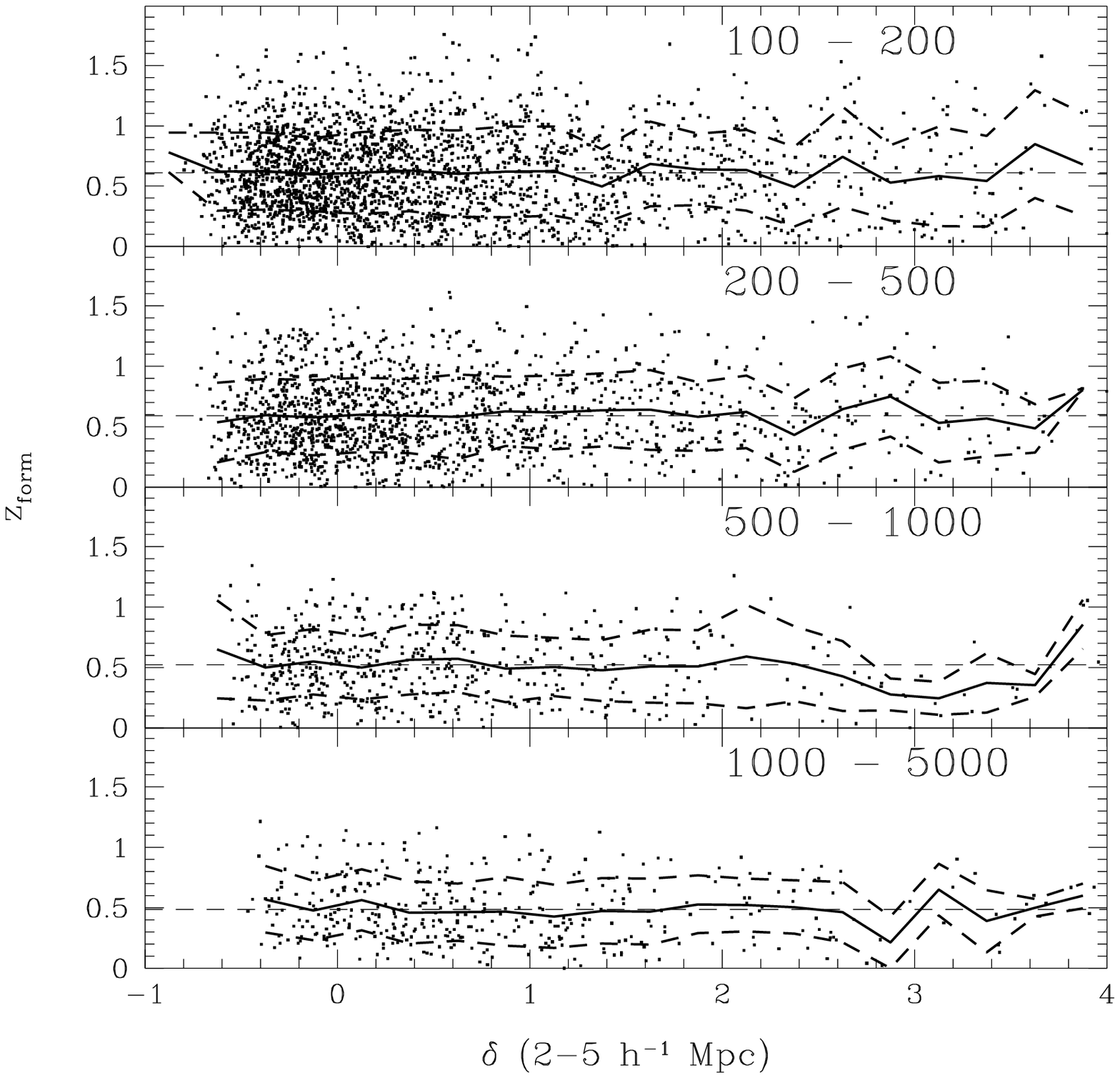}
}
\caption{\label{fig3}
The formation redshift is plotted against overdensity in the shell between 2
and 5 h$^{-1}$ Mpc for
halos in different mass ranges. For an explanation of the lines, see figure 2.}
\end {figure}

Figures 2 and 3 demonstrate that the {\em average} spin and formation time of
halos are not correlated with
their environments. The next step is to demonstrate that the full probability
distributions
are also independent of local overdensity. This is shown in figure 4, where we
plot $P(\lambda)$
and $P(z_{form})$, for three different ranges in overdensity $\delta$. As can
be seen, the
curves are almost indistinguishable from each other. Again, there is perhaps a
slight shift
towards larger spins in overdense regions.
The distribution of  $\lambda$ is consistent with that found in earlier work
(e.g.
Barnes \& Efstathiou 1987).

\begin{figure}
\centerline{
\epsfxsize=14cm \epsfbox{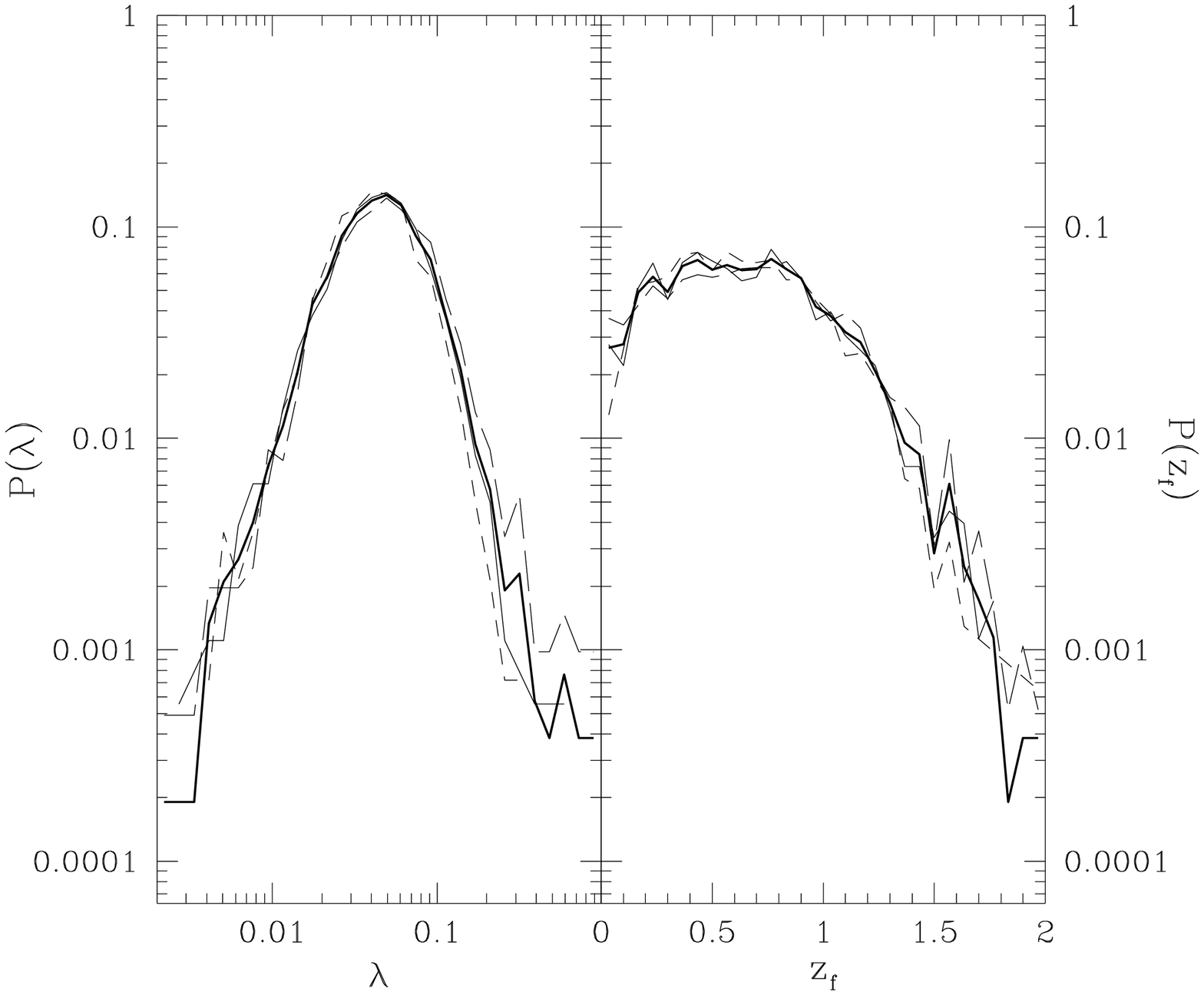}
}
\caption{\label{fig4}
The probability distribution of the spins and the formation redshifts of halos
in regions of different overdensity $\delta$.
$\delta$ is calculated in
a 10 h$^{-1}$ Mpc sphere for the left-hand panel, and in a 2-5 h$^{-1}$ Mpc
shell
for the right-hand panel. The thick solid line
in both panels represents the full probability distribution.
In the left panel,
the short dashed, thin solid, and long-dashed lines are for $-1 < \delta <
-0.1$,
$-0.1 < \delta < 0.3$, and $0.3 < \delta$.
In the right panel,
the short dashed, thin solid, and long-dashed lines are for $-1 < \delta <
-0.1$,
$-0.1 < \delta < 0.6$, and $0.6 < \delta$.
Bins in $\delta$ have been chosen so that they contain an equal number of
haloes.}
\end {figure}

Finally, figure 5 is a mosaic illustrating a further subset of the
possibilities we have
explored.
In the two left panels, we show the concentration index $c_{10}$  versus the
overdensity
evaluated in spheres of radius 10 h$^{-1}$ Mpc surrounding each halo, and
the shape axis ratio $p_1/p_3$ versus the tidal field axis ratio
($\phi_{,11}/\phi_{,33}$).
In the two right panels, we plot
the spin parameter and the formation redshift versus principal axes ratios of
the
tidal field and the density shear field respectively. (Note that the apparent
gaps
in the distribution of $ \lambda$ and $p_{1}/p_{3}$ at small positive values of
$\phi_{,11}/\phi_{,33}$  result from the ordering of the eigenvalues and
the values that the $\phi_{,ii}$ assume in practice.) Once again, no
correlations are found, either in the mean values of the halo quantities, or in
their
full probability distributions. It is interesting that neither the spins, nor
the
distribution of shapes of dark halos,
depend on the surrounding tidal field.

\begin{figure}
\centerline{
\epsfxsize=14cm \epsfbox{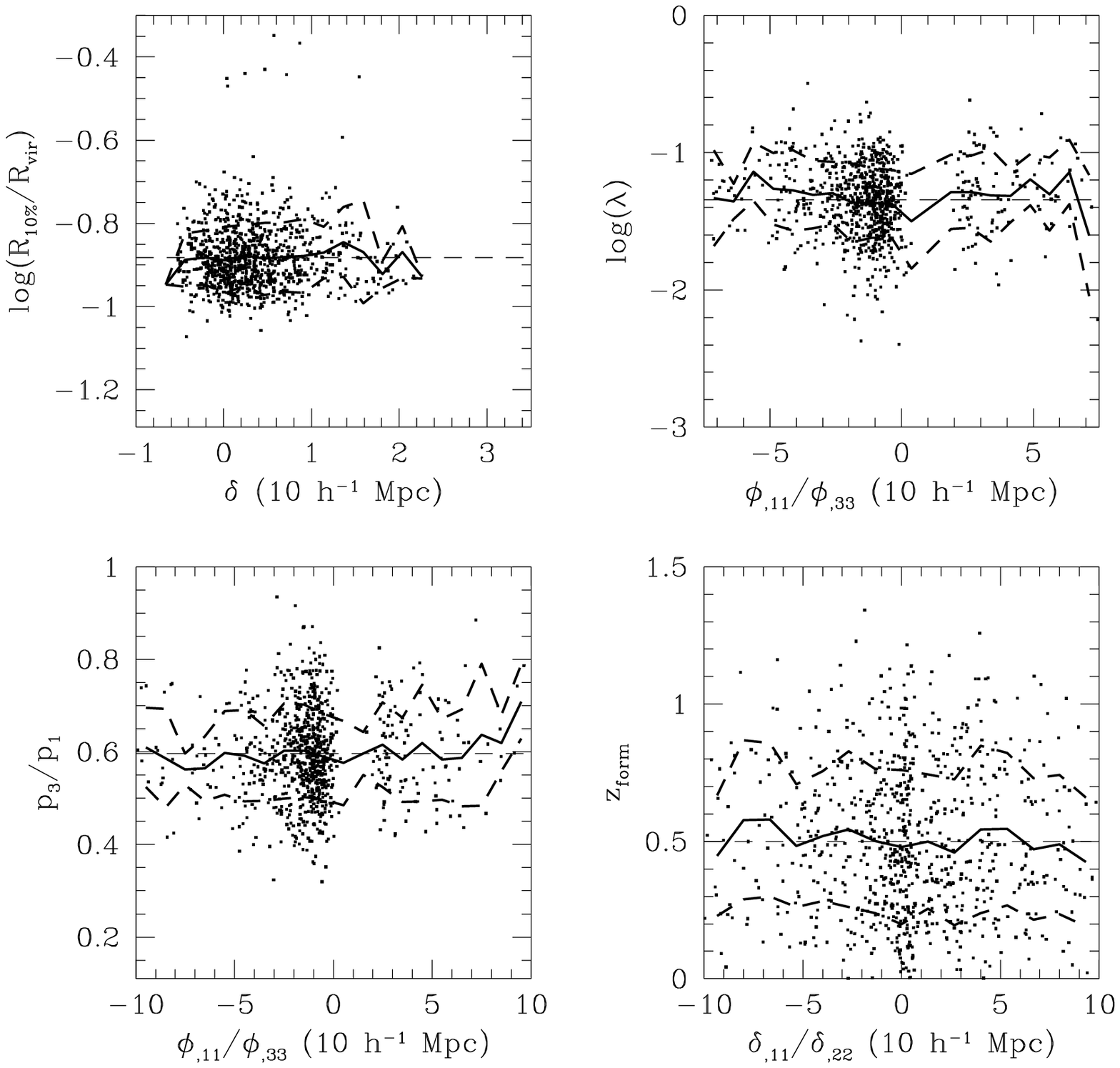}
}
\caption{\label{fig5}
A mosaic of four  different correlations between halo properties and
environment. The top left panel
shows the concentration index ($c_{10}= R_{10}/R_{vir}$) as a function of
$\delta$ in a sphere.
The bottom left panel shows the relationship between the shape axis ratio
$p_3/p_1$
and the the ratio of the first and third eigenvalues of the tidal field.
In the top right panel, the spin parameter is plotted against the
ratio of the first and third eigenvalues of the tidal field. In the bottom
right panel,
formation redshift is plotted against the ratio of the first and second
eigenvalues of the
density shear field. Only halos in the mass range 500-5000 are shown.  }
\end {figure}

\section {Discussion and Conclusions}

In this paper, we have demonstrated that mass is the only halo property that
correlates
significantly with
local environment. The dependence of
the halo
mass function on environment can be understood in terms of a simple extension
of the Press-Schechter theory.

It is  interesting to note that the fact that halo formation histories do
not depend on local overdensity, is a
prediction of the extended Press-Schechter theory in the excursion set
formulation of
Bond et al (1991). In this approach, the standard formulae are obtained by
assuming
that the mass of a halo to which any given mass element belongs, can be
followed
by studying the behaviour of the initial linear density field as it is smoothed
with
a succession of sharp k-space filters. In this model, the history of each mass
element
of a halo of mass $M$ (and thus the formation history of the halo itself) is
statistically independent of the future of the element and thus of the halo's
environment.
Navarro, Frenk \& White (1996) have demonstrated that
 the density profile of dark matter
halos in CDM-like cosmologies  has a ``universal'' form, and that the
characteristic density
of a halo is related simply to its formation
time. It is thus also not
surprising that our concentration parameters $c_{10}$ and $c_{20}$  turn
out to be independent
of environment.
There has also been substantial analytic work concerned with the
angular momentum
generated by tidal torques acting on local
density
maxima in the linear density field. It has been argued (Blumenthal et al 1984;
Hoffman 1986)
that the height of a peak is anticorrelated with its angular momentum, on the
basis
that high peaks collapse early, so there is not much time for tidal torques to
act.
If this was the case,  overdense regions of the Universe, in which high peaks
are
more probable, should harbor halos with systematically lower angular momentum.
Heavens \& Peacock (1988) pointed out  that this effect is
counterbalanced
by the fact that higher peaks experience greater tidal torques. For realistic
power
spectra, they estimate that the two effects should nearly cancel out. Steinmetz
\& Bartelmann (1995)
obtain a similar result by modelling the formation of a halo as a collapse of a
homogeneous
ellipsoid acted upon by tidal shear from the surrounding matter. Both analyses
are
consistent with the very weak environment-dependence of spin in our data.
The shapes of dark matter halos have proven more difficult to calculate using
analytic methods.
Dubinski (1992) has investigated halo shapes in high-resolution N-body
simulations and finds
that there is no relationship between the shape of the initial density peak and
the shape
of the final collapsed halo. On the other hand, West, Villumsen \& Dekel (1991)
and Tormen (1997) have demonstrated a clear tendency for the major axis of
cluster-sized halos to align with larger-scale structures.

The main astrophysical implication of our conclusion is that clustering
differences between galaxies
of different types must arise purely because different galaxies sample
different mass halos.
Semi-analytic models of galaxy formation in hierarchical cosmologies
demonstrate
that it is possible to explain  many of the clustering trends seen in
the data in this way (Kauffmann, Nusser \& Steinmetz 1997). In these models,
elliptical galaxies
form when disk galaxies of comparable mass merge with each other. These mergers
occur preferentially
in groups at redshifts $\simeq 1$. The groups then coalesce to form clusters
and superclusters.
As shown by Kauffmann, White \& Guiderdoni (1993) and Baugh, Cole \&
Frenk(1996), the
fraction of
ellipticals increases strongly  with halo mass in this picture and ellipticals
thus end up
more clustered than spirals.
In addition, the luminosity of the central galaxy in a halo
scales in proportion to the halo mass,  simply because more gas is able
to cool and form
stars in more massive halos. Luminosity-dependent clustering is thus a natural
outcome
of the models. Finally, it should also be noted that galaxies are assumed to
be stripped of their surrounding dark matter
once they have been accreted by a larger system, such as a group or
cluster. They thus
lose their supply of new cold gas, their star formation rates decline, and
their stellar
populations redden and fade. Red, gas-poor galaxies are thus found
predominantly in
high mass halos and are again predicted to be more clustered.
The semi-analytic models at present provide no explanation for the clustering
differences between
galaxies of different surface brightnesses. One might speculate that low
surface-brightness disks
are more fragile and are more easily destroyed in high-density
environments. More detailed modelling
is necessary, however,  before a conclusion
can be drawn as to whether this will work in practice.

\vspace{0.8cm}

\large
{\bf Acknowledgments}\\
\normalsize
We thank Simon White, Adi Nusser and Joe Silk for helpful discussions.

\pagebreak
\Large
\begin {center} {\bf References} \\
\end {center}
\normalsize
\parindent -7mm
\parskip 3mm

Barnes, J. \& Efstathiou, G., 1987, ApJ, 319, 575

Baugh, C.M., Cole, S. \& Frenk, C.S., 1996, MNRAS, 283, 1361

Benoist, C., Maurogordato, S., Dacosta, L.N., Cappi, A. \& Schaeffer, R., ApJ,
1996
472, 452

Blumenthal, G.R., Faber, S.M., Primack, J.R. \& Rees, M.J., 1984, Nature, 311,
517

Bond, J.R., Cole, S., Efstathiou, G. \& Kaiser, N. 1991, ApJ, 379, 440

Bothun, G.D., Schombert, J.M., Impey, C.D., Sprayberry, D. \& McGaugh, S.S.,
1993, AJ, 106, 530

Bower, R.J., 1991, MNRAS, 248, 332

Cayette, V., Kotanyl, C., Balkowski, C. \& Van Gorkom, J.H., 1994, AJ, 107,
1003

Cole, S., Arag\'on-Salamanca, A., Frenk, C.S., Navarro, J.F.
\& Zepf, S.E., 1994, MNRAS, 271, 781

Couchman, H.M.P., Thomas, P.A. \& Pearce, F.R., 1995, ApJ, 452, 797

Dalcanton, J.J., Spergel, D.N. \& Summers, F.J., 1997, ApJ, 482, 659

Davis, M. \& Geller, M.J., 1976, ApJ, 208, 13

Dressler, A., 1980, ApJ, 237,351

Dubinski, J., 1992, ApJ, 401,441

Heavens, A. \& Peacock, J., 1988, MNRAS, 232, 339

Hoffman, Y., 1986, ApJ, 301, 65

Kauffmann, G., White, S.D.M. \& Guiderdoni, B. 1993, MNRAS, 264, 201 (KWG)

Kauffmann, G., Nusser, A. \& Steinmetz, M., 1997, MNRAS, 286, 795

Kauffmann, G., Diaferio, A.,  Colberg, J. \& White, S.D.M., 1997, in
preparation

Kennicutt, R.C. 1983, AJ, 88, 483

Lacey, C., Guiderdono, B., Rocca-Volmerange, B. \& Silk, J., 1993, ApJ, 402, 15

Lacey, C. \& Cole, S., 1993, MNRAS, 262, 627

Lacey, C. \& Cole, S., 1994, MNRAS, 271, 676

Loveday, J., Maddox, S.J., Efstathiou, G. \& Peterson, B.A., 1995, ApJ, 442,457

Mo, H.J., Einasto, M., Xia, X.Y. \& Deng, Z.G., 1992, MNRAS, 255, 382

Mo, H.J., McGaugh, S.S. \& Bothun, G.D., 1994, MNRAS, 267, 129

Mo, H.J. \& White, S.D.M., 1996, MNRAS, 282, 347

Mo, H.J., Mao, S. \& White, S.D.M., 1997, MNRAS, submitted

Navarro, J.F., Frenk, C.S. \& White, S.D.M., 1996, ApJ, 462, 563

Park, C.B., Vogely, M.S., Geller, M.J. \& Huchra, J.P., 1994, APJ, 431, 569

Peacock, J.A. \& Dodds, S.J., 1994, MNRAS, 267, 1020

Pearce, F.R. \& Couchman, H.M.P., 1997, New Astronomy, submitted

Steinmetz, M. \& Bartelmann, M., 1995, MNRAS, 272, 570

Tormen, G., 1997, MNRAS, 290, 411

Valotto, C.A. \& Lambas, D.G., 1997, ApJ, 481, 594

West, M.J., Villumsen, J.V. \& Dekel, A., 1991, ApJ, 369, 287

White, M., Gelmini, G. \& Silk, J., 1995, Phys. Rev. D., 51, 2669

White, S.D.M. \& Rees, M.J., 1978, MNRAS, 183, 341

White, S.D.M. \& Frenk, C.S., 1991, ApJ, 379, 52

White, S.D.M., Efstathiou, G. \& Frenk, C.S.,1993, MNRAS, 262, 1023
\end {document}